\documentclass[fleqn,usenatbib]{mnras}

\usepackage[T1]{fontenc}

\DeclareRobustCommand{\VAN}[3]{#2}
\let\VANthebibliography\thebibliography
\def\thebibliography{\DeclareRobustCommand{\VAN}[3]{##3}\VANthebibliography}

\usepackage{graphicx}	
\usepackage{amsmath}	
\usepackage{amssymb}	
\usepackage{multirow}

\usepackage{newtxtext,newtxmath}

\usepackage{xcolor} 
\usepackage{tikz} 
\usetikzlibrary{patterns}
\usetikzlibrary{positioning}
\usetikzlibrary{quotes,angles}
\usetikzlibrary{intersections}

\usepackage{soul} 

\graphicspath{{Figures/}}

\title[{\tt RELAGN}: AGN SEDs and Black Hole Spin]{Estimating Black Hole Spin from AGN SED Fitting: The Impact of General-Relativistic Ray Tracing}

\author[S. Hagen and C. Done]{
Scott Hagen,$^{1}$\thanks{E-mail: scott.hagen@durham.ac.uk}\thanks{GitHub: \url{https://github.com/scotthgn/RELAGN}}
Chris Done$^{1}$
\\
$^{1}$Centre for Extragalactic Astronomy, Department of Physics, University of Durham, South Road, Durham DH1 3LE, UK\\
}

\date{Accepted XXX. Received YYY; in original form ZZZ}

\pubyear{2023}

\newcommand{\rout}{r_{\mathrm{out}}}
\newcommand{\risco}{r_{\mathrm{isco}}}
\newcommand{\mdot}{\dot{m}}
\newcommand{\Mdot}{\dot{M}}
\newcommand{\Mdedd}{\dot{M}_{\mathrm{Edd}}}
\newcommand{\Mbh}{M_{\mathrm{BH}}}
\newcommand{\fcol}{f_{\mathrm{col}}}
\newcommand{\ktw}{kT_{e, w}}
\newcommand{\kth}{kT_{e, h}}
\newcommand{\Msol}{M_{\odot}}
\newcommand{\Ledd}{L_{\mathrm{Edd}}}

\setstcolor{red}

\begin{document}
\label{firstpage}
\pagerange{\pageref{firstpage}--\pageref{lastpage}}
\maketitle

\defcitealias{Kubota18}{KD18}

\begin{abstract}

Accretion disc model fitting to optical/UV quasar spectra requires that the highest mass black holes have the highest spin, with implications on the hierarchical growth of supermassive black holes and their host galaxies over cosmic time. 
However, these accretion disc models did not include the effects of relativistic ray tracing. Here we show that gravitational redshift 
cancels out most of the increase in temperature and luminosity from the smaller radii characteristic of high spin. Disc models which include the self consistent general relativistic ray tracing do not fit the UV spectra of the most massive quasars ($\log M/\Msol \geq 9.5$), most likely showing that the disc structure is very different to that assumed. We extend the relativistic ray tracing on more complex disc models, where the emission is not limited to (colour temperature corrected) black body radiation but can instead be emitted as warm and hot Comptonisation. We demonstrate this on the broadband (UV/X-ray) spectrum of Fairall 9, a local intensively monitored 'bare' AGN (no significant intrinsic cold or warm absorption). We show that including relativistic corrections does make a difference even to these more complex models, but caution that the inferred black hole spin depends on the assumed nature and geometry of the accretion flow. 
Additionally, we make our model code publicly available, and name it {\sc relagn}.

\end{abstract}




\section{Introduction}

Accretion discs around black holes powering Active Galactic Nuclei (AGN) are generally understood in the framework of the \citet{Shakura73} disc model. This describes an optically thick, geometrically thin structure with constant mass accretion rate as a function of radius. The resulting luminosity and temperature of the disc increases with decreasing radius, until close to the inner edge of the disc, where the stress free inner boundary condition starts to affect the emissivity. This gives an overall spectral energy distribution (SED) which is a sum of blackbodies from all the disc annuli. 

The importance of general relativistic corrections to the original Newtonian framework was soon recognised, 
firstly in setting the inner disc radius at the innermost stable circular orbit ($\risco$), secondly in giving corrections to the disc emissivity \citep{Novikov73}, and thirdly in sculpting the observed spectrum at infinity from ray tracing the fast orbital velocities of the disc material through the curved spacetime of strong gravity \citep{Cunningham75}. All these effects  are incorporated in several publicly available models for the disc emission, tailored for stellar mass black hole binaries (e.g. {\sc kerrbb} and {\sc bhspec}: \citealt{Li05, Davis05} in {\sc xspec}). 
Fully relativistic disc models were also developed for the lower temperatures (UV/EUV) expected for AGN discs (e.g. \citealt{Sun89, Laor89, Laor90, Hubeny01}). 
However, there are no corresponding models in the public domain  
which are widely used to fit broadband AGN spectra in the optical/UV. The well known {\sc QSOFIT/pyQSOFIT} \citep{Shen11, Guo18} use a power law approximation for the AGN disc continua, as does the similar {\sc QSFit} code \citep{Calderone17} as both focus on 
disentangling the intrinsic emission from the reprocessed lines and recombination radiation. Models which instead focus on separating out the AGN from the host galaxy use fixed template AGN spectra
(see e.g the compilation in Fig.\,1 of 
\citealt{Thorne21}).

The use of these phenomenological models is driven in 
in part by the realisation that AGN spectra are intrinsically more complex than a simple sum of black-body spectral models. Early work focused on electron scattering, leading to modified black-body emission from the photosphere \citep{Shakura73, Czerny87}. The true absorption opacity $\kappa_\nu(T)$ can be much less than the electron scattering opacity, $\kappa_T$, leading to an effective photosphere which extends deeper into the disc (where the temperature is higher) at higher frequencies (e.g. \citealt{Ross92, Hubeny01}). The emission from a single radius can be approximated as a colour temperature corrected black-body, $B_\nu(\fcol T)/\fcol^4$ where $\fcol \sim 1.4-2$ \citep{Shimura95, Davis05, Slone12, Done12}.
This puts a characteristic bend in the UV spectrum, as there is no significant contribution from electron scattering at temperatures below $\sim 10^4$~K where hydrogen becomes neutral
(e.g. \citealt{Czerny87}). This bend is observed in Quasar spectra (e.g \citealt{Zheng97, Telfer02, Shang05, Barger10, Shull12, Lawrence12}). UV line driven disc winds also become important at similar temperatures \citep{Laor14}, 
and mass losses from the disc will also redden the UV continuum shape \citep{Slone12}. 

Irrespective of the exact shape of the disc emission, these models do not predict spectra that extend far into the soft X-rays, yet there is a ubiquitous 'soft X-ray excess' component which seems to point back to the UV downturn as well as the separate X-ray tail from a hot Comptonising corona (e.g \citealt{Laor97, Porquet04, Gierlinski04}). The soft X-ray excess can be modelled by an additional Comptonising plasma which is warm, and quite optically thick \citep{Mehdipour11, Done12, Mehdipour15, Kubota18, Petrucci18}. The large optical depth means that this component is likely the disc itself, perhaps indicating a change in the vertical disc structure such that the emission does not completely thermalise \citep{Rozanska15,Jiang20} as would otherwise be expected if the energy were dissipated mainly in the midplane as in standard disc models. 

Thus there seem to be three separate components required to fit the observed broadband SED, namely a disc, warm Comptonisation for the soft X-ray excess and hot Comptonisation for the X-ray tail. It is very difficult to robustly fit three unconstrained components to the data, especially as there is an unavoidable gap in coverage from at least  10--200~eV due to the combined effects of gas absorption and dust reddening through our Galaxy (e.g. \citealt{jin09}). Instead, 
recent progress has stressed that the entire broadband AGN spectra can be modelled by radial stratification of the accretion flow. The emissivity is assumed to still be Novikov-Thorne, as appropriate for a disc, but this power is emitted as either black-body, warm Comptonisation or as hot Comptonisation depending on radius (\citealt{Done12}, updated in \citealt{Kubota18}; hereafter \citetalias{Kubota18}). The three components are then tied together by energetics, so give robust fits to the data.

These models 
are very successful in fitting the optical/UV/X-ray spectra of samples of AGN (\citealt{Jin12a, Jin12b, Collinson15, Mitchell23}) as well as  describing more detailed spectra of individual objects \citep{Matzeu16, Done16, Czerny16, Hagino16, Hagino17, Porquet18}. However, while the model is based on the fully relativistic Novikov-Thorne emissivity, it does not include ray tracing from the disc to the observer. This can have a significant effect, especially at low inclinations where there is little projected blueshift from the disc motion to compensate for gravitational and transverse redshift, and for high spins; where all the relativistic effects are stronger. \citet{Done13} showed an approximate way to incorporate these corrections using existing well known relativistic smearing models incorporated into {\sc xspec}. 
\citet{Porquet19} showed that these do make a significant difference in the spin derived from fitting broadband data to a bare (probably face on) AGN, Akn120. More recently \citet{Dovciak22} developed a broad-band SED model for AGN that does take all relativistic effects into account; namely {\sc kynsed}. However, this only includes radial stratification between a disc and hot Comptonisation region,
so does not model the soft X-ray excess, though it does also include relativistic effects on the illuminated disc reprocessed emission. 

Here we use the relativistic transfer functions of \citet{Dovciak04} to properly include the relativistic
ray tracing as a function of radius on the {\sc agnsed} 
emitted spectrum i.e. including the warm Comptonisation region as well as the disc and corona. 
We demonstrate the new code, {\sc relagn}, on the AGN Fairall 9, and make it publicly available for {\sc xspec} \citep{Arnaud96} and as a stand-alone {\sc python} module. We also demonstrate the importance of using these relativistic corrections for fitting optical/UV spectra of the highest mass Quasars, and show the impact this has on the high spins determined for these objects by \citet{Capellupo15, Capellupo16}. 

Other potential uses are that this new code 
allows a disc geometry with arbitrary truncation radius to be tested on data. This is especially important in black hole binaries to directly test truncated disc/hot inner flow models. All current public relativistic ray traced disc codes (see above)
are hardwired with a disc extending down to the innermost stable circular orbit (ISCO). Instead, {\sc relagn} allows the disc to be smoothly modelled from an inner radius at the ISCO where full relativistic corrections are required, to truncation far from the black hole where these effects are small.

The new code is also the first to explicitly include ray tracing on the soft X-ray excess warm Comptonised disc, which are currently popular in the literature. This component is energetically important, and often dominates the AGN SED. GR ray tracing should then be important in determining the observed spectrum, and our code is so far the only one which incorporates this.

Additionally, we also put ray tracing on the {\sc qsosed} model. This is a version of {\sc agnsed} which hardwires all the geometry parameters to follow the observed trends in SED shape as a function of $L/\Ledd$ \citepalias{Kubota18}. Thus the model predicts an entire broadband SED from the physical parameters of 
mass, mass accretion rate and black hole spin (see e.g. \citealt{Mitchell23}).
We name this {\sc relqso}, and include this in the public release. 

This paper is organised as follows. In section 2 we give details on modelling the SED, starting with the spectrum emitted in the comoving disc frame (which we also refer to as the rest frame) in section 2.1, and then incorporating the relativistic ray tracing in section 2.2. We then provide example models in sections 2.3 (standard disc) and 2.4 (full broad-band SED), with model cavaets for the hot corona in section 2.5. In section 3 we apply {\sc relagn} to data, starting with a colour-temperature corrected accretion disc fit to Optical/UV X-shooter data on one of the highest mass, lowest mass accretion rate quasars (SDSS J092714.49+000400.9) in section 3.1, before moving on to a full broad-band SED fit (Optical/UV to X-ray) of a more typical local Quasar, Fairall 9, in section 3.2. 

\section{Modelling the SED}

When modelling the SED we start from the emitted spectrum in the comoving disc frame, before applying the relativistic transfer functions to obtain the SED seen by a distant observer. The comoving calculations are identical to those described in \citetalias{Kubota18} for the {\sc agnsed} model. For completeness we give a brief description of this  model first, before including the relativistic transfer functions. Throughout we will use the standard notation for radii, where $R$ denotes the radius in physical units and $r$ is the radius in dimensionless gravitational units. These are related through $R = r R_{G}$, where $R_{G} = G\Mbh/c^{2}$. Similarly, for mass accretion rate, $\Mdot$, denotes the physical mass accretion rate in g/s, while $\mdot$ is dimensionless. 
These are related through the Eddington rate, such that $\Mdot = \mdot \Mdedd$, defined including the spin dependent efficiency $\eta(a_*)$ so that 
$L_{\mathrm{Edd}} = \eta(a_*) \Mdedd c^{2}$. 

\subsection{The {\sc AGNSED} model - Comoving Calculations}

\begin{figure}
    \centering
    \begin{tikzpicture}
        
        \draw[pattern=crosshatch, pattern color=blue] (0, 0) circle (30pt) ;
        \filldraw[color=white] (0, 0) circle (10pt) ;
        \filldraw[color=black] (0, 0) circle (5pt) ;
        
        \filldraw[color=green] (-1.05, 0.07) -- (-2.85, 0.2) -- (-2.85, -0.2) -- (-1.05, -0.07) ;
        \filldraw[color=red] (-2.85, 0.2) -- (-4.65, 0.33) -- (-4.65, -0.33) -- (-2.85, -0.2);
        
        \node (risco) at (-0.15, 1.5) {\large{$\risco$}} ;
        \draw[very thick] (-0.38, 0) -- (-0.38, 1.3) ;
        
        \node (rh) at (-1, 1.5) {\large{$r_{h}$}} ;
        \draw[very thick] (-1.05, 0) -- (-1.05, 1.3) ;
        
        \node (rw) at (-2.8, 1.5) {\large{$r_{w}$}} ;
        \draw[very thick] (-2.85, 0) -- (-2.85, 1.3) ;
        
        \node (rout) at (-4.5, 1.5) {\large{$\rout$}} ;
        \draw[very thick] (-4.65, 0) -- (-4.65, 1.3) ;
    
    \end{tikzpicture}
    \caption{The {\sc agnsed} model geometry, consisting of a standard disc (red) from $\rout$ to $r_{w}$, at which point it fails to thermalise and enters the warm Comptonising region (green). This continues down to $r_{h}$, after which the disc evaporates into the hot Comptonising region (blue), which extends down to $\risco$}
    \label{fig:mod_geom}
\end{figure}
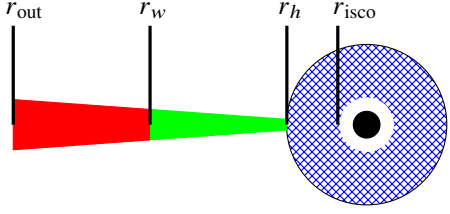

As in {\sc agnsed} we divide the accretion flow into three components: the standard disc, the warm Comptonising region, and the hot Comptonising region. This geometry is sketched in Fig.\,\ref{fig:mod_geom}.

Assuming \citet{Novikov73} emissivity, we have a radial temperature profile that goes as $T^{4}_{NT}(R) \propto R^{-3} f(R)$, where $f(R)$ describes the radial disc structure in the Kerr metric \citep{Page74}. The accretion flow is then split into annuli of width $\Delta R$ and temperature $T_{NT}(R)$. For the outer standard disc we assume all the emission thermalises, such that each annulus emits like a black-body $B_{\nu}(T_{NT}(R))$ with luminosity given by $2 \times 2\pi R \Delta R \sigma T^{4}_{NT}(R)$. Here the extra factor of $2$ is due to the disc emitting from both sides, and $\sigma$ is the Stefan-Boltzmann constant.

We extend the {\sc agnsed} model to include the possibility of a colour temperature correction 
on the standard disc region. This can either be set to a specified value, or switched to the 
predicted value for each annulus temperature assuming the vertical structure of a standard disc 
\citep{Davis05}. This has $\fcol=1$ below 
$\sim 10^4$~K, where hydrogen ionisation is mostly neutral so there are few free electrons for scattering, then it increases rapidly to $ > 2$, then declines to $\sim 1.7$. 
(see \citealt{Done12}). This differs from
\citetalias{Kubota18}, who hardwired the colour temperature correction at unity as they assumed this was subsumed in the warm Comptonisation region (see below). 

However, a colour-temperature correction is 
just an approximation to a more complex spectrum. We consider also the case where the disc fails to thermalise even to a colour temperature corrected black-body,
forming instead a warm Comptonisation region \citep{Petrucci13, Petrucci18} below some radius $r_{w}$ (green region in Fig.\,\ref{fig:mod_geom}).
This could form if the dissipation 
region moves higher into the photosphere than expected in a standard disc, forming a slab geometry above an underlying dense disc structure. 

Following \citetalias{Kubota18} we tie the seed photons in the warm Comptonisation region to the underlying disc, such that the seed photon temperature is simply $T_{NT}(R)$, and each annulus has luminosity $4 \pi R \Delta R \sigma T_{NT}^{4}(R)$ (i.e. no colour temperature correction). We can then calculate the Comptonised spectrum from each annulus using {\sc nthcomp} \citep{Zdziarski96, Zycki99}, and assuming that each annulus emits seed photons like a black-body. The disc photosphere is optically thick, with $\tau >> 1$, and has a covering fraction of unity in our assumed geometry, so all the black-body seed photons are Comptonised \citep{Petrucci18}.
We also assume that the entire warm Comptonisation region has a constant electron temperature, $\ktw$, and photon index, $\Gamma_{w}$, and leave these as parameters in the {\sc relagn} code.

The warm Comptonised region extends down to a radius $r_{h}$, below which it evaporates into a hot, optically thin, geometrically thick flow forming the X-ray tail (see Fig.\,\ref{fig:mod_geom})
\citep{Narayan95, Liu99, Rozanska00a}. Again, following \citetalias{Kubota18}, we consider the dominant process in this region to be Compton scattering and that the geometry can be approximated as a sphere surrounding the black hole; with inner radius $\risco$. The total power produced by this region will then be a sum of the power dissipated in the accretion flow between $r_{h}$ and $\risco$, $L_{h, \mathrm{diss}}$, and the power from the disc photons intercepted by the corona, $L_{h, \mathrm{seed}}$ (see \citetalias{Kubota18} for a detailed description on calculating these). Since the dominant mechanism is Compton scattering, we again use {\sc nthcomp} to calculate the spectral shape, and normalise to the total luminosity $L_{h} = L_{h, \mathrm{diss}} + L_{h, \mathrm{seed}}$. Like the warm Comptonising region, we consider the electron temperature within the corona, $\kth$, to be constant and leave it as a parameter in the code. For the seed photon temperature, $kT_{\mathrm{seed}, h}$, we firstly assume that these are dominated by photons from the inner edge of the disc. If this emits as warm Comptonisation then these will have a temperature given by $T_{NT}(R_{h}) \exp(y_{w})$, where $y_{w}$ is the Compton y-parameter for the warm Comptonisation region \citepalias{Kubota18}. Otherwise they will simply have temperature $T_{NT}(R_{h})$. 
We assume these seed photons form a black-body distribution, and we leave the photon index, $\Gamma_{h}$, as a free parameter.

We stress here that when we calculate the hot Comptonised emission we only take into account the power balance through the flow (i.e energy conservation). We do not simultaneously consider conservation of photon number, though Compton scattering should include this. The hardest AGN spectra, those close to the 'changing state' transition at $\log{\mdot} \lesssim -1.6$ may have an issue with being photon starved in a similar way to the low/hard state in black hole binaries (see e.g. \citealt{Poutanen18}), but otherwise AGN have spectra where there are copious UV/far UV seed photons so an additional seed photon source such as cyclo-synchrotron (see e.g. \citealt{Malzac09,vrum09}) is not required.

Finally, the total rest frame SED is the sum of the contribution from each annulus. The flow is split into geometrically spaced radial bins, which are chosen such that each annulus is confined to a single region; i.e $\rout$, $r_{w}$, $r_{h}$, and $\risco$ are treated as explicit bin edges.

Unlike {\sc agnsed} we do not consider the re-processing of the hot corona flux when it illuminates the outer disc/warm corona regions. This is because the illumination pattern depends on the unknown vertical structure of the corona, and the reprocessed spectrum depends on the unknown density (and hence ionisation) structure of the illuminated disc/warm Compton region. \citetalias{Kubota18} assume that all the flux from a uniformly emitting spherical hot corona thermalises in order to maximise the reprocessing, but even this makes only a small (less than 10\%) increase in the UV flux except for the hardest AGN spectra. Additionally, recent 
intensive monitoring campaigns show that the majority of the observed lagged UV re-processed signal originates from large scale height material at large distances (the BLR or a wind on the inner edge of the BLR) rather than the disc itself (e.g \citealt{Mehdipour16, Dehghanian19b, Chelouche19, Kara21}).

\subsection{{\sc RELAGN} - the SED seen by a distant observer}

The rest frame SED is affected by both special and general relativistic effects, due to the fast orbital motion and strong gravity close to the black hole  \citep{Cunningham75,Fabian89, Chen89}. 
A common method for incorporating relativistic effects is to convolve the intrinsic spectrum (often just a narrow gaussian line) with a relativistic transfer function, using one of the many available models; e.g {\sc kdblur}, {\sc kyconv}, {\sc relconv},  etc \citep{Laor91,  Dovciak04, Garcia14, Dauser14}. Here, the transfer functions encode all the relevant relativistic effects, and can be used to  determine the emission seen by a distant observer. These can in principle be applied to continuum spectra; simply convolving our rest frame SED with one of the transfer functions. However, this is not the case where the continuum changes shape with radius as in our models. Each disc annulus produces a different spectrum, and is subject to slightly different relativistic effects. This means we need to apply the transfer functions to each annulus separately, before adding up their contributions to the total SED.

\begin{figure*}
    \centering
    \includegraphics[width=\textwidth]{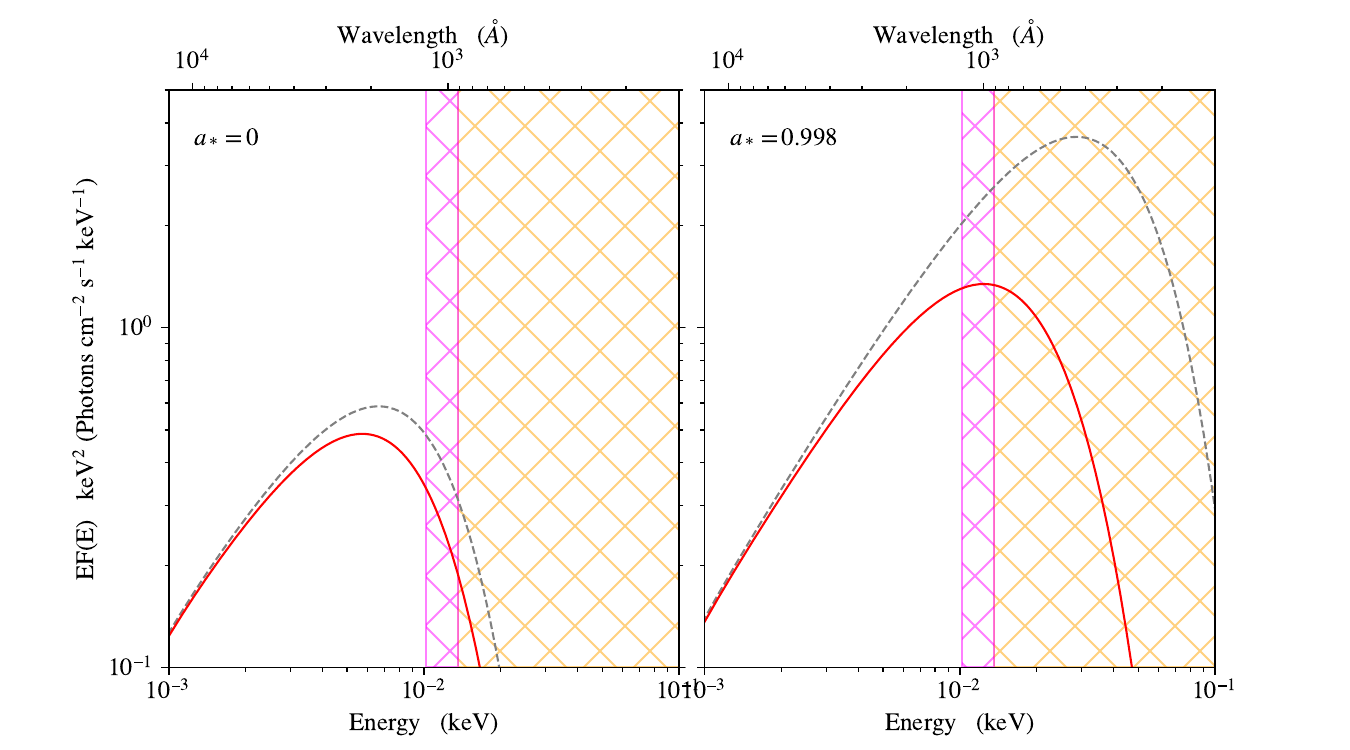}
    \caption{A comparison of relativistic (solid red lines) and non-relativistic (dashed-grey lines) SEDs for a standard disc extending from $\log \rout = 3$ to $\risco$, in the non-rotating (left) and maximally spinning (right) cases. These have been calculated for a $\Mbh = 10^{9}\,\Msol$ black hole, with physical mass accretion rate $\Mdot = 10^{26}$\,g/s, and observed at an inclination of $\cos(i) = 0.9$. The orange hatch indicates the unobservable region (beyond the Hydrogen threshold of $\sim 13.6$\,eV), while the magenta hatch region indicates where Lyman-$\alpha$ emission becomes important ($\sim 10.2$\,eV).}
    \label{fig:relVnonrelSED_disc}
\end{figure*}

\citet{Done13} approximated this by using the tabulated relativistic transfer functions for each emission region separately, as also used by \citet{Porquet19}. However, this is only an approximation as the spectrum emitted from each region changes shape as a function of radius. Additionally the tabulated transfer functions are normalised to unity, i.e. they redistribute photon energy but conserve photon rate. This is not accurate: photons in the inner disc are time dilated, and lightbending means that many of these are deflected away from the observers line of sight. Here instead we do the transfer from each annulus explicitly, avoiding approximations, and use the intrinsic normalisation of the transfer functions. 
We use the {\sc kynconv} transfer functions \citep{Dovciak04} as these have a parameter switching the normalisation to intrinsic (unlike {\sc kdblur}: \citealt{Laor91} which does not include a switch, or 
{\sc relconv}: \citealt{Dauser14} where the code has to be recompiled with the switch set outside of {\sc xspec} as an environment variable).

To calculate the SED seen by a distant observer we first calculate the comoving emission from each annulus, following {\sc agnsed} \citepalias{Kubota18}, as described in the previous section. This is then convolved with {\sc kyconv}, to give the annular emission seen by a distant observer, before we add each contribution together to create the total SED. We work with an internal radial grid which is sufficiently fine that the emissivity is approximately constant across an annulus (see Appendix A. for details). Since our code calculates the comoving emissivity profile internally we set the emissivity indexes in {\sc kyconv} to $0$ (i.e {\sc kyconv} is forced to assume constant emission across an annulus). 

The disc transfer functions are clearly appropriate for the outer standard disc and warm Compton regions, but only approximately capture the general relativistic effects on the hot flow (see Fig. \ref{fig:mod_geom}). We sketch this as a spherical region, but hot flow is more likely to have density and dissipation concentrated towards the midplane (see e.g. \citealt{liska22}). 
Nevertheless, it should be rotating at close to Keplerian, and have emissivity peaking close to the black hole, so the 
disc transfer functions give an estimate of the expected general relativistic effects.

\subsection{Example model spectra - the UV disc}

\begin{figure*}
    \centering
    \includegraphics[width=\textwidth]{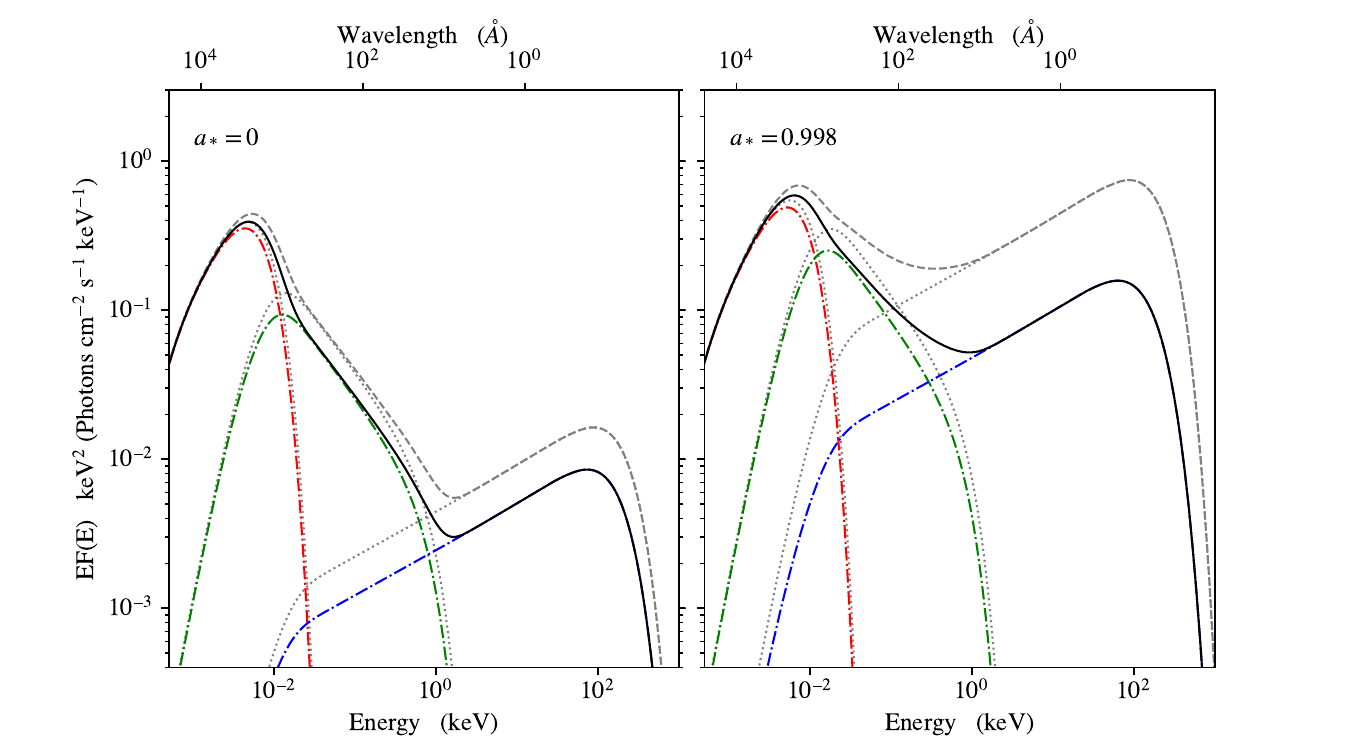}
    \caption{A comparison of relativistic (solid and coloured lines) and non-relativistic (dashed/dotted grey lines) for the full SED model, for both a non-rotating (left) and maximally spinning (right) black hole. As in Fig. \ref{fig:relVnonrelSED_disc} these have been calculated for a $\Mbh = 10^{9}\,\Msol$ black hole, accreting at $\Mdot = 10^{26}$\,g/s, and observed at an inclination of $\cos(i) = 0.9$. Here the model consists of a standard outer disc (red) extending from $\rout = 10^{3}$ to $r_{w} = 20$, a warm Comptonisation region (green) from $r_{w} = 20$ to $r_{h} = 10$, and a hot Comptonisation region (blue) from $r_{h} = 10$ to $\risco$.}
    \label{fig:relVnonrelSED_full}
\end{figure*}

In this section we highlight the differences between {\sc agnsed} and {\sc relagn} using a fixed black hole mass of $10^9M_\odot$ accreting at $\dot{M}=10^{26}$~g/s (in physical units), viewed at close to face on ($\cos{i}=0.9$). This mass accretion rate corresponds to $\log L/\Ledd =-1.4$ for spin 0, and $-0.7$ for spin 0.998. We set $\fcol=1$ in all the plots in this section, since for these high mass black holes we do not expect the disc temperature to become sufficiently high for $\fcol$ to become important (see \citealt{Done12} for details on $\fcol$).

We start with the simple case where the accretion flow consists only of a standard disc extending from $\rout = 10^{3}$ down to $\risco$, shown in Fig.\,\ref{fig:relVnonrelSED_disc}. Higher spin means $\risco$ is smaller so there is additional higher temperature/higher luminosity emission from these smaller radii. However, these are also the radii which are most affected by the strong special and general relativistic effects (fastest orbital velocity and strongest gravitational redshift). At low inclination gravitational redshift dominates over the Doppler red/blueshifts, so much of the additional hot emission is lost to the observer. This was clear in early disc models (e.g. \citealt{Sun89}), but fits of these models to data were hindered by lack of black hole mass estimates, leaving this crucial parameter free (e.g. \citealt{Laor97}). 

For all but the highest mass, lowest luminosity AGN, the optical/UV continuum from the disc is a power law to a good approximation as the disc peak is in the EUV region which is challenging ($>10$~eV: magenta) or impossible ($>13.6$~eV: orange) to observe directly. However, we now have single epoch mass estimates from line width/continuum relations \citep{Vestergaard02, Vestergaard06}, so all broad line AGN have a default black hole mass. We also now have some very high mass quasars, where the peak temperature predicted by the disc models give a rollover close to the observable UV, and these are typically at redshift $z>1$, making the peak more visible. 
\citet{Mitchell23} used the {\sc relagn} code to show that fits to composite SDSS spectra of the highest mass, lowest luminosity (hence lowest predicted peak disc temperature) quasars are significantly affected by the relativistic ray tracing. The reasons for this become clear in Fig.\,\ref{fig:relVnonrelSED_disc}. At $a_*=0$
(left panel) the disc spectrum of a high mass black hole peaks within the observable window, effectively allowing us to see the emission from the inner-most orbit; which is most strongly affected by relativistic effects. The grey dashed line shows the spectrum without the ray tracing, while the solid red line shows the effect of including this on the disc emission. 
The right panel shows the same physical mass accretion rate onto a maximally spinning black hole ($a_*=0.998$). The intrinsic spectrum (grey dashed line) is much brighter and hotter, peaking in the unobservable region. However, this does not mean that the GR ray tracing effects are likewise hidden. Including the ray tracing (red solid line) strongly redshifts the hottest emission, predicting that the turn over is still observable. This is simply because as spin increases, $\risco$ moves closer in to the black hole, and so any relativistic effects must in turn become stronger. What we show here, is that although increasing spin will significantly boost the emission in the rest frame, the increased strength of the relativistic effects compensates for some of this (depending on inclination). This will have a significant impact for spin estimates of the highest mass Quasars, as we will show in section 3.1.

The black hole mass used for the example in Fig.\,\ref{fig:relVnonrelSED_disc} was intentionally chosen to highlight the
observable impact of the GR ray tracing on the Optical/UV spectrum of accretion discs. This would not be so obvious for a lower mass black hole. It is clear in Fig.\,\ref{fig:relVnonrelSED_disc} that the impact from GR ray tracing dissipates as we move to lower energies in the spectrum, and the fully relativistic spectrum eventually approaches the non-relativistic case for sufficiently low energy. This is due to the low energy emission originating at larger radii in the disc, where the special and general relativistic effects are all considerably smaller. For a low mass black hole ($\lesssim 10^{7}\,\Msol$ for the same Eddington ratios considered here) the SED peaks in the unobservable region, even for the non-spinning case, so there is minimal impact on the observed SED.

\subsection{Example model spectra - the full broad-band SED}

The same concept as in the previous subsection applies when we consider a more complex SED, shown in Fig.\,\ref{fig:relVnonrelSED_full}. We set $r_w=20$ and $r_h=10$, but this means that the hot Compton region extends from $r=10-6$ for $a_*=0$, but from $10-1.23$ for maximal spin so it is much more luminous in the high spin case. For both spins, the hot Compton emission is the 
part which originates in the innermost regions of the flow, so this will be most strongly affected by the relativistic corrections. The normalisation of this component is obviously suppressed at (close to) face on inclination, and the high energy rollover is also redshifted. 

The soft Compton emission is from larger radii, so the ray tracing effects are less marked. Nevertheless, they are still present, with a clear suppression in the normalisation 
which is similar for both spins as 
this component extends over a fixed radial range of $r= 20-10$ for this example and so the GR ray tracing effects are similar, irrespective of spin. However, there is still a clear difference in the intrinsic luminosity of this component with spin, due to the stress free inner-boundary condition suppressing the emissivity below $r=10$ for low spin but not at high spin. 
Hence the luminosity of this component for these radii is somewhat dependent on spin, but this is due to the intrinsic (rest frame) emission rather than any difference in GR ray tracing.

\subsection{The Effects of GR on the Observed Coronal Flux}

In Fig. \ref{fig:relVnonrelSED_full} it is clear that the GR ray tracing has a significant effect on the observed X-ray power, even though the corona extends to $r_{h} = 10$ for this example. This is because the intrinsic dissipation in the hot flow is assumed to follow the Novikov-Thorne emissivity, so extends inwards from $r_{h}$ to $\risco$. Thus the emissivity weighted mean radius for the hot coronal emission is less than $r_{h}$, and decreases substantially for increasing spin as the peak of the emissivity moves inwards, so general relativistic effects are stronger.
This contrasts with the often used 'lamppost' corona geometry, where the dissipation region is compact, so all the emission is produced at a single radius.

\begin{figure}
    \centering
    \includegraphics[width=\columnwidth]{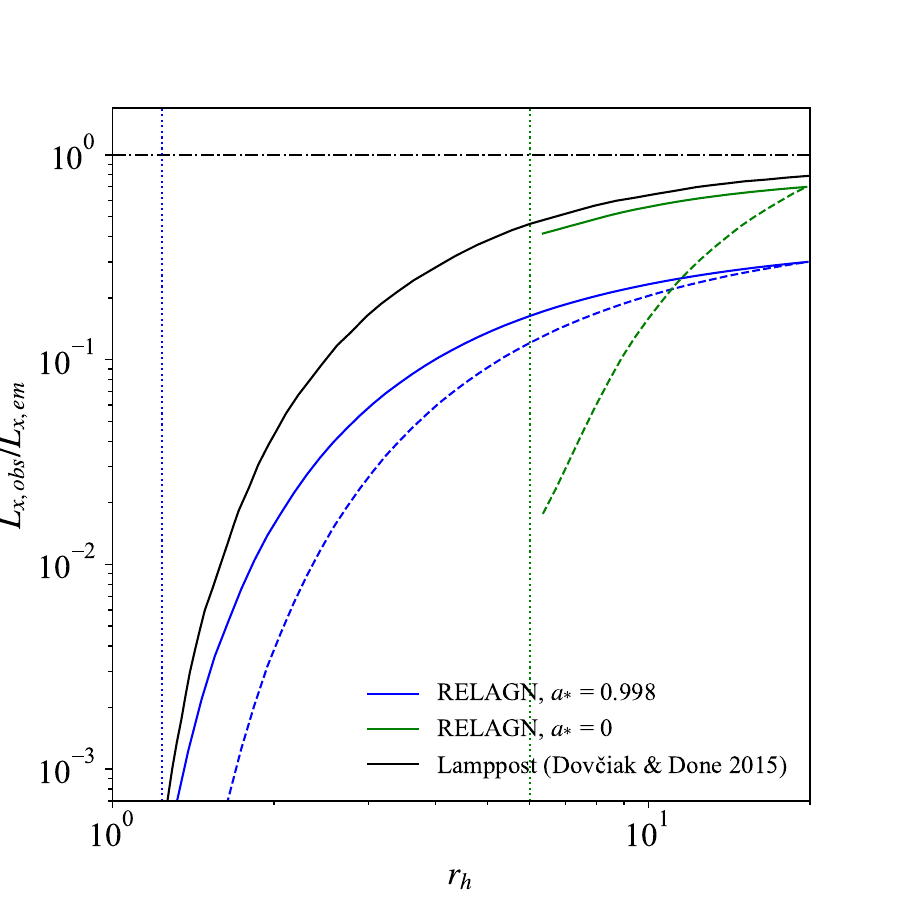}
    \caption{The green (spin 0) and blue (maximal spin) solid lines show the ratio of observed (at $i=30^\circ$) to emitted X-ray flux from the {\sc relagn} model for different hot corona radii, $r_h$. General relativistic effects (predominantly gravitational redshift) increase as $r_h$ decreases, suppressing the observed flux. The black solid line shows this ratio for a lamppost corona at height $r_h$ above a maximally spinning black hole for comparison (taken from \citealt{Dovciak15}), showing a similar suppression of the observed to emitted flux. However, there is a difference in that the intrinsic lamppost flux is assumed constant as a function of $r_h$, whereas the {\sc relagn} model has smaller $L_{x, \mathrm{em}}$ for smaller $r_h$. The green and blue dashed lines show the total dimming (intrinsic and relativistic effects) relative to that emitted at $r_h=20$ (i.e $L_{x, \mathrm{obs}}/L_{x, \mathrm{em}}(r_{h}=20)$). The vertical coloured lines show $\risco$.
    }
    \label{fig:lxObs_v_Lem}
\end{figure}

We explore the effect of ray tracing on the observed emission from an accretion powered X-ray hot region viewed at $30^\circ$ in more detail in 
Fig. \ref{fig:lxObs_v_Lem}. This shows the 
ratio of observed to emitted luminosity from the X-ray hot region for 
decreasing $r_{h}$ for a black hole of maximal (blue solid line) and zero (green solid line) spin. This integrates over the emission from the hot flow seen at infinity, so includes both the drop in normalisation of the power law section of the hot Compton spectrum as well as the redshifted temperature
(see Fig. \ref{fig:relVnonrelSED_full}). 

Similarly to the disc spectra discussed in Section 2.3, increasing the black hole spin increases the intrinsic luminosity due to the increased efficiency from the flow extending closer to the black hole, but much of this extra emission is redshifted and/or lost down the black hole rather than escaping to infinity. e.g. in Fig. \ref{fig:relVnonrelSED_full} the comoving X-ray power is $\sim 50\times$ larger for $r_h=10$ for maximal spin compared to spin zero, while the power seen at infinity is only $15\times$ larger. We also note that the X-ray hot accretion flow can be less efficient than the thin disc expectation of Novikov-Thorne, reducing the X-ray flux still further.

The black solid line instead shows the same plot for a lamppost corona for maximal spin. The accretion powered hot flow always has stronger dimming than the lamppost as the mean radius at which the emission is produced is smaller than $r_h$, but the two geometries merge for lamppost height close to $\risco$ for a maximally spinning black hole.
However, there is a key intrinsic difference between the hot Comptonisation region models and a lamppost. The lamppost has 
no direct connection to the energy generating accretion process, so 
moving the lamppost height makes no difference in its intrinsic flux. Conversely, in our model, reducing $r_{h}$ directly reduces the accretion power available to heat the hot corona, so that the X-ray source is intrinsically dimmer as well as subject to larger relativistic effects. The green and blue dashed lines show the total dimming for an accretion powered X-ray source relative to its luminosity at $r_h=20$ (i.e $L_{x, \mathrm{obs}}/L_{x, \mathrm{em}}(r_{h}=20)$), and shows a much faster decrease in observed X-ray luminosity.

Fig. \ref{fig:lxObs_v_Lem} shows that 
any high spin model where the emission arises from less than $2.5R_g$ has less than 10\% of the intrinsic X-ray power reaching infinity
(see also e.g. \citealt{niedzwiecki16,Dovciak15}). In fact, even for a low spin black hole, if $r_{h}$ is sufficiently close to $\risco$ the observed X-ray emission will be minimal. This also means that some of our simplifying assumptions about the corona structure are appropriate as the data do not enter the regime where gravitational effects are extreme, since 
the predicted X-ray emission would be too dim 
in the case where $r_{h}$ is sufficiently small for the entire corona to experience the most extreme gravitational effects (see Fig. \ref{fig:lxObs_v_Lem}).

\section{Application to data}

\subsection{UV disc spectra: J0927+0004}

\begin{figure*}
    \centering
    \includegraphics[width=\textwidth]{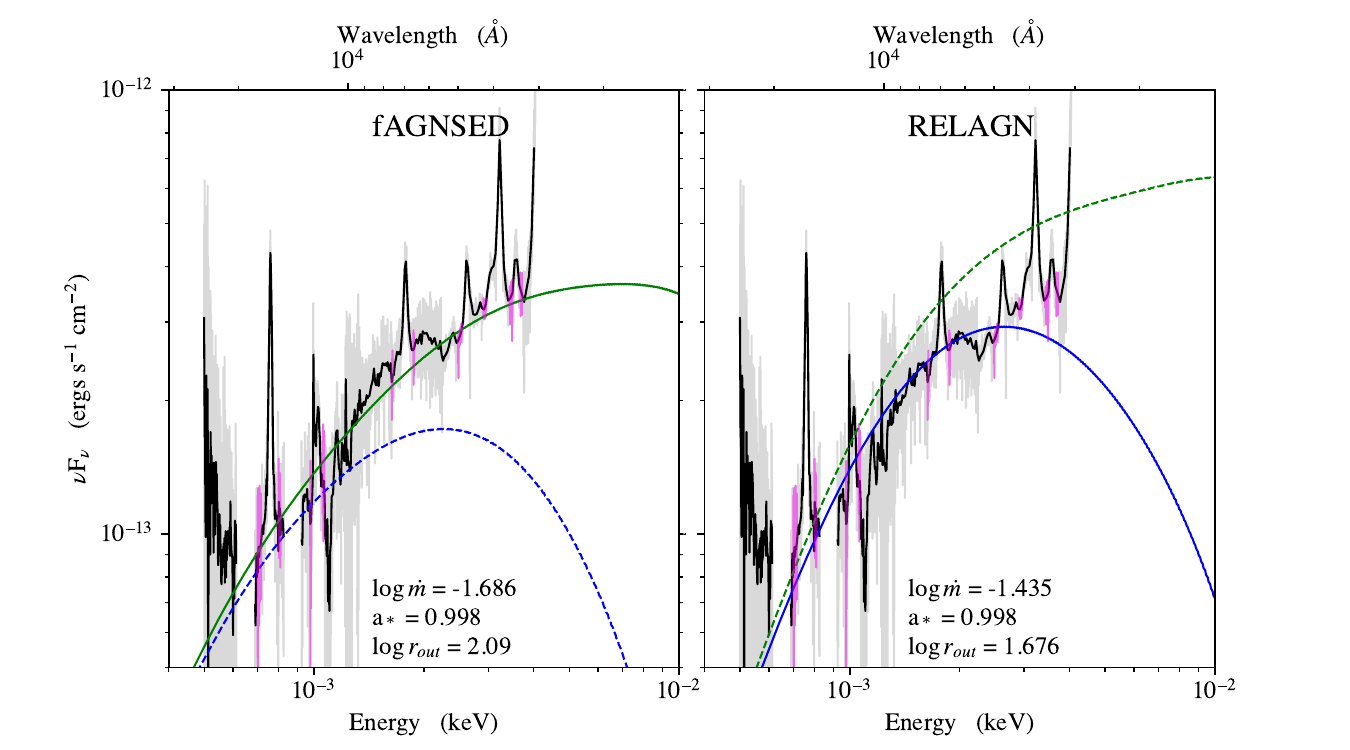}
    \caption{Fits to X-shooter data on J0927+0004, using f{\sc agnsed} (left, green) and {\sc relagn} (right, blue). The data are de-reddened by \citet{Fawcett22}, and shown by the solid black line, which has been smoothed for clarity. We also show the un-smoothed data as the opaque grey line. The line-free continuum windows used for fitting are shown in magenta. In both panels we show the alternative model for the same parameters. i.e in the left panel the dashed blue line is {\sc relagn}  using fit parameters from f{\sc agnsed}, while in the right panel the dashed green line is f{\sc agnsed} using fit parameters from {\sc relagn}. This is to highlight how GR ray tracing will take a seemingly acceptable fit and make it invalid. It is clear that although the non-relativistic treatment is able to fit the data by forcing maximal spin, once GR ray tracing is taken into account the standard disc model is not able to fit the data; instead predicting a turn-over in the spectrum well below the data.}
    \label{fig:xshooter}
\end{figure*}

We first explore the impact of the GR ray tracing on results from fitting pure disc models to the optical/UV spectra of Quasars. The best examples of these are from the sample of \citet{Capellupo15} with X-shooter data. The GR ray tracing affects the spectrum most around the disc peak, so we pick the object with the highest mass, lowest mass accretion rate in this sample so that it has the lowest predicted peak disc temperature, so that it can be studied in the observable UV region. This selects
SDSS J092715.49+000400.9 (hereafter J0927+0004). These data are publically available at the X-shooter archive. However, here we used the de-reddened and binned data from \citet{Fawcett22} 
(V. Fawcett, private communication), where they were used as part of their control sample of un-reddened quasars. J0927+0004 has a high black hole mass, $\log \Mbh/\Msol =9.2-9.3$ low Eddington ratio, $\log \mdot = -1.4\,-\,-1.3$, and minimal intrinsic reddening \citep{Mejia-Restrepo16, Capellupo16, Fawcett22}. This, combined with an SDSS (DR7) redshift of $z=1.4845$ \citep{Schneider10} means that the disc SED peaks within the observable UV/Optical bandpass. 

We fit the data with the {\sc relagn} code
with $r_w=r_h=\risco$ (so it only produces disc emission) and set the colour temperature correction to that of \citet{Done12}. 
We fix the black hole mass to $\log \Mbh/\Msol = 9.2$ \citep{Fawcett22}, and the inclination to $\cos(i) = 0.87$. 
This leaves only three free parameters: spin ($a_{*}$), mass-accretion rate ($\mdot$), and outer radius ($\rout$).

Fig.\,\ref{fig:xshooter}a shows fits to 'line-free' continuum regions (purple, defined as in \citealt{Capellupo15}) with the GR ray tracing turned off (solid green line). We call this f{\sc agnsed}, as it is the same as {\sc agnsed} but with colour temperature corrections included (though these are not important here as the disc temperature is mostly below $10^4$~K: see also \citealt{Mitchell23}). This fit without the GR ray tracing reproduces the results from \citet{Capellupo15, Capellupo16} and \citet{Fawcett22}, all of whom use the 
standard disc models of \citet{Slone12} (which include the colour temperature correction, and Novikov-Thorne emissivity, but not ray tracing) to estimate black hole parameters. 
This gives a good fit for $a_*=0.998, \log L/L_{Edd}=-1.69$ and 
$\log \rout = 2.09$ which is within a factor two of the expected self-gravity radius ($r_{sg} \sim 200$ for $\log \Mbh/\Msol = 9.2$ and $\log \mdot = -1.69$; \citealt{Laor89}). 

The blue dashed line in Fig.\,\ref{fig:xshooter}a shows the effect of turning on the GR ray tracing for this model. 
Plainly the data are now not at all well fit at the highest UV energies as photons from the inner regions are highly redshifted. 
The solid blue line in Fig.\,\ref{fig:xshooter}b 
shows instead a fit with {\sc relagn}, including the GR ray tracing. Clearly the fit is much worse than in Fig.\,\ref{fig:xshooter}a, as the model never convincingly fits the highest energy UV emission. The model has shifted to higher $\log L/\Ledd$ but this is not sufficient to recover a high enough temperature peak even with maximal spin. The green dashed line shows the corresponding model with GR ray tracing removed.

Clearly the disc model is not an adequate description of these data once GR ray tracing is taken into account. This is consistent with the results of \citet{Mitchell23}, who fit {\sc relagn} to stacked SDSS spectra, and showed that the disc model fails once the relativistic ray tracing is included for their high mass bins ($\log \Mbh/\Msol > 9.5$). This shows that this is a common issue for both spectra of individual objects and stacked samples of high mass quasars. 
\citet{Mitchell23} also tried to fit with the warm Comptonisation component extending over the entire disc, and found that it could fit but required that parameters change in mass and Eddington ratio in a way that appears fine tuned (their Fig. 14).

\begin{figure*}
    \centering
    \includegraphics[width=\textwidth]{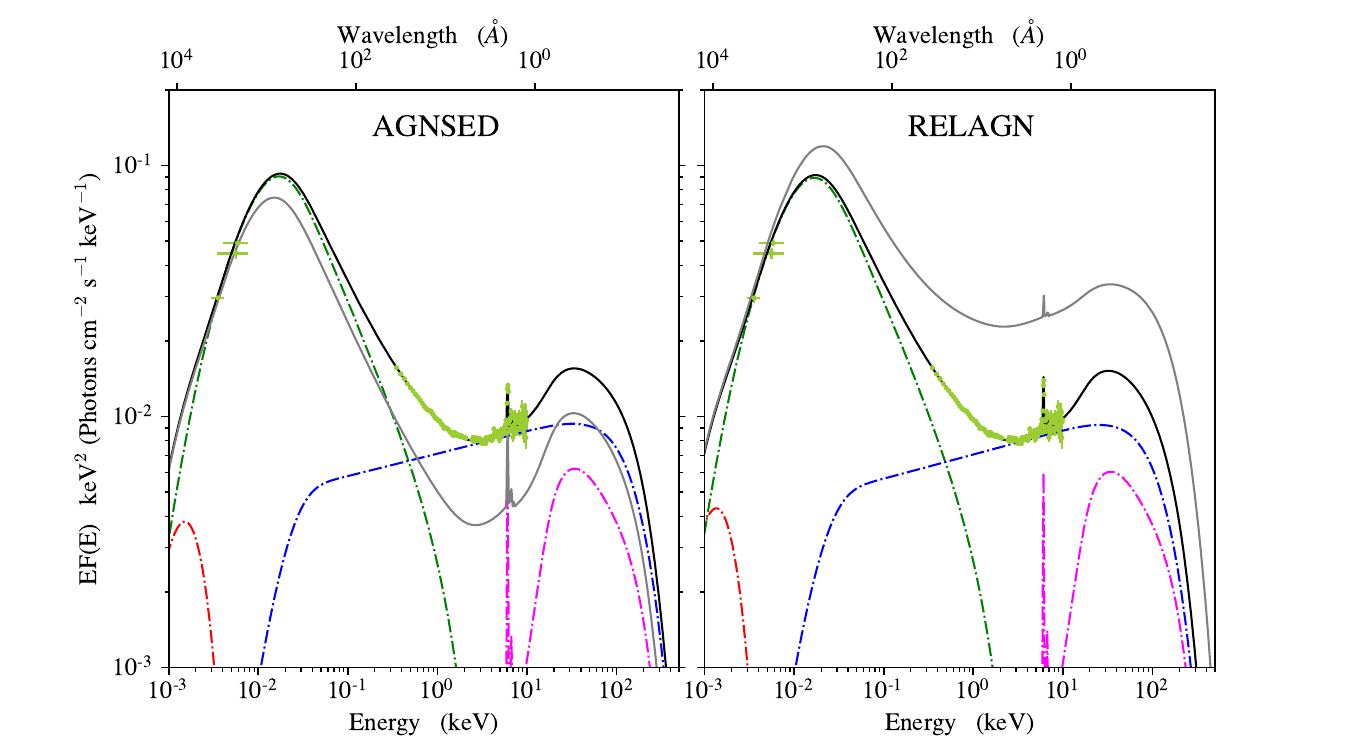}
    \caption{Fits to archival data on Fairall 9, using {\sc agnsed} (left) and {\sc relagn} (right). The magenta dashed-dotted line indicates the reflection component, included to model the iron line in the X-ray spectrum. The other coloured dashed-dotted lines correspond to the hot Comptonised component (blue), warm Comptonised component (green) and standard disc component (red). In each panel there is over-plotted a solid grey line, which shows the alternative model for the same parameters. i.e in the left panel the grey line is {\sc relagn} using the fit parameters from the {\sc agnsed} fit, and vice-versa for the right panel.}
    \label{fig:F9_SED}
\end{figure*}

This impacts on our understanding of black hole accretion and growth. 
\citet{Capellupo15, Capellupo16} 
used their disc fits without the GR ray tracing to show that higher mass black holes have higher spins. This favours models in which the
accretion flow has a preferential angular momentum direction over a prolonged period, whereas low-spin favours chaotic accretion where the disc angular momentum is more or less random \citep{King08, Dotti13}. Hence, \citet{Capellupo15, Capellupo16} suggested that their results were consistent with a spin-up model of black hole evolution; where the spin increases as mass increases (i.e the mildly anisotropic accretion scenario). 
Instead, we show that GR has a significant impact on the model spectrum, such that the simple disc models are not a good description of the data even at maximal spin. Without a good model fit we cannot reliably estimate black hole spin, and hence cannot draw conclusions on the nature of black hole spin evolution. We consider more complex models of the accretion flow emission below.

\subsection{Broad band SED: Fairall 9}

Black hole spin estimates rely on probing the emission from the innermost stable circular orbit, $\risco$. Studies based on the observed Optical/UV alone have to assume that a disc model is an adequate description of the data, and that it extends down to $\risco$. In the previous subsection we saw that the inclusion of relativistic ray tracing breaks these assumptions for the most massive quasars. However, our full model is able to predict the broad-band SED, extending from Optical/UV to the X-ray. In the context of our model it is also assumed that the X-rays originate in the innermost regions of the flow, and hence could act as a probe for $\risco$. Therefore, we now apply our new {\sc relagn} model to broad-band spectral data on Fairall 9. 

The UV/Optical data are the galaxy subtracted time-averaged UVOT data from the 1st year observation campaign of Fairall 9 by \citet{Hernandez20}, and were converted to an {\sc xspec} readable count-rate using the conversion factors in \citet{Poole08}.
In an ideal world we would use the simultaneous 
Swift-XRT and/or NICER spectra to define the soft and hard X-ray flux. However, as discussed in Appendix D of \citet{Hagen23}, there are several issues with these data (potential pile up with Swift-XRT, background estimation for NICER). Instead, we use an archival XMM-Newton observation \citep{Lohfink16} which has a similar UV flux level, so probably represents a similar state. Note that although the Swift-UVOT data are not simultaneous with the XMM observations, the Swift-UVOT UVW2 filter has a larger effective area, providing better count statistics.

We follow \citet{Hagen23} and fit the data with 
a global photoelectric absorption component, {\sc phabs}, to account for galactic absorption, as well as a reflection component, {\sc pexmon} \citep{Nandra07, Magdziarz95}, to model the Fe-K$\alpha$ line. To account for any smearing in the Fe-K$\alpha$ line we convolve {\sc pexmon} with {\sc rdblur} \citep{Fabian89}. The total {\sc xspec} model is then: {\sc phabs * (main + rdblur*pexmon)}, where {\sc main} is used to denote either {\sc agnsed} or {\sc relagn}.

Fig.\,\ref{fig:F9_SED}a shows the data (corrected for the X-ray absorption and deconvolved from the response) fit with 
{\sc agnsed} as in \citet{Hagen23} (black solid line). This provides a reasonable fit to the data, but once we apply GR ray tracing to this fit (grey solid line) it becomes clear that 
this fit under-predicts the X-ray power, even though the best fit {\sc agnsed} model had  only moderate spin of $a_{*} \sim 0.7$. 

Fig.\,\ref{fig:F9_SED}b shows the data fit with 
{\sc relagn} (solid black line). Here we can still get an acceptable fit (unlike the pure disc models for the highest mass quasars in the previous subsection), but for higher Eddington ratio (with $\log \mdot = -0.980$ compared to $\log \mdot = -1.215$ for {\sc agnsed}) and a higher black hole spin ($a_{*} = 0.938$ compared to $a_{*} = 0.715$ for {\sc agnsed}).
The increase in Eddington ratio and spin is compensating for the reduction in observed power from the GR ray tracing (grey line shows the rest frame emission). Best fit parameters for each model are given in Table \ref{tab:SED_pars}.

Thus the complex emission accretion flow models are able to constrain black hole spin from the energetics. However, we caution that in 
this particular case the X-ray data are not simultaneous. More generally there are still caveats as it is the innermost parts of the accretion flow which are most sensitive to spin, which here is the X-ray corona. We have incorporated GR ray tracing on this component assuming a disc geometry and velocity field, but this is not consistent with the schematic (Fig.\,\ref{fig:mod_geom}) which envisages a larger scale height (hence sub-Keplarian rotation) flow. This may not even follow the Novikov-Thorne dissipation if 
e.g. there is magnetic connection across $\risco$ or if the flow truncates at larger radii due to torques from misalignment with the black hole spin and especially if some part of this accretion power is used to produce the radio jet. 
Additionally, we assume here that the UV data are dominated by the disc continuum, however 
the intensive continuum reverberation campaigns are revealing that there is a substantial component in the UV which arises from 
re-processing in a wind on the inner edge of the BLR \citep{Mehdipour16, Dehghanian19b, Chelouche19, Kara21, Netzer22, Hagen23}. Hence, we encourage the reader to use spin estimates from {\sc relagn} with caution, as a guide to the system energetics, rather than a perfect description of the accretion flow.

{\renewcommand{\arraystretch}{1.6} 
\begin{table}
    \centering
    \begin{tabular}{c|c|c|c}
         & & {\sc AGNSED} & {\sc RELAGN} \\
         \hline
         \multicolumn{4}{|c|}{- - - - - - - - - - - {\sc phabs} - - - - - - - - - - -} \\
        $N_{H}$ & $10^{20}$\,cm$^{-2}$ &  3.5 & 3.5 \\
        \multicolumn{4}{|c|}{- - - - - - - - - - - {\sc main} - - - - - - - - - - - } \\
        Mass & $M_{\odot}$ & $2 \times 10^{8}$ & $2 \times 10^{8}$ \\
        Distance & Mpc & $200$ & $200$ \\
        $\log \mdot$ & $\log(\Mdot/\Mdedd)$ & $-1.215^{+0.012}_{-0.010}$& $-0.980^{+0.020}_{-0.041}$ \\
        Spin & & $0.715^{+0.052}_{-0.057}$ & $0.938^{+0.016}_{-0.052}$ \\
        $\cos(i)$ & & 0.9 & 0.9 \\
        $kT_{e, h}$ & keV & 100 & 100 \\
        $kT_{e, w}$ & keV & $0.383^{+0.057}_{-0.033}$ & $0.416^{+0.050}_{-0.045}$ \\
        $\Gamma_{h}$ & & $1.912^{+0.021}_{-0.029}$ & $1.906^{+0.021}_{-0.031}$ \\
        $\Gamma_{w}$ & & $2.824^{+0.028}_{-0.026}$ & $2.817^{+0.015}_{-0.033}$ \\
        $r_{h}$ & & $9.32^{+0.91}_{-0.91}$ & $8.9^{+1.2}_{-0.4}$ \\
        $r_{w}$ & & $324^{+158}_{-108}$ & $325^{+127}_{-138}$ \\
        $r_{\text{out}}^{\dagger}$ & & $=r_{sg}$ & $=r_{sg}$ \\
        $f_{\text{col}}^{\dagger \dagger}$ & & 1 & 1 \\
        $h_{\text{max}}^{\dagger \dagger \dagger}$ & & 10 & 10 \\
        Redshift & & 0.045 & 0.045 \\
        \multicolumn{4}{|c|}{- - - - - - - - - - - {\sc rdblur} - - - - - - - - - - - } \\
        Index & & -3 & -3 \\
        $r_{\text{in}}$ & & $265.78^{+339}_{-113}$ & $442^{+5959}_{-166}$ \\
        $r_{\text{out}}$ & & $10^{6}$ & $10^{6}$ \\
        Inc & deg & 25 & 25 \\
        \multicolumn{4}{|c|}{- - - - - - - - - - - {\sc pexmon} - - - - - - - - - - - } \\
        $\Gamma$ & & $=\Gamma_{h}$ & $=\Gamma_{h}$ \\
        $E_{c}$ & keV & $10^{3}$ & $10^{3}$ \\
        Norm & $\times 10^{-3}$ & $4.51^{+0.64}_{-0.62}$ & $4.27^{0.56}_{-0.81}$ \\
        \hline
        $\chi^{2}_{\nu}$ & 165 d.o.f & 1.363 & 1.351 \\
        \hline
     \end{tabular}
    \caption{Best fit parameters of {\sc agnsed} and {\sc relagn} to the data on Fairall 9 (Fig.\,\ref{fig:F9_SED}). The parameters in {\sc main} belong to {\sc agnsed}/{\sc relagn} (depending on the column). Values with no errors were left frozen in the fitting process.
    \newline
    $\dagger$: We fix $\rout$ to the self-gravity radius, $r_{sg}$, from \citet{Laor89}
    \newline
    $\dagger \dagger$: $f_{\mathrm{col}}$ is fixed to 1 in {\sc agnsed}.
    \newline
    $\dagger \dagger \dagger$: $h_{\text{max}}$ will only affect the contribution of the seed photons to $L_{h}$, as we have left re-processing off in {\sc agnsed} (and is neglected in {\sc relagn}).}
    \label{tab:SED_pars}
\end{table}
}

\section{Summary and Conclusions}

We have developed a fully relativistic version of {\sc agnsed}, referred to as {\sc relagn}. This incorporates general relativisitc ray tracing as well as the classic relativisitic Novikov-Thorne disc emissivity, while allowing the spectra emitted 
from each radius to be more complex than just a (colour temperature corrected) black-body. 
Including the ray tracing has a significant impact on the predicted SED, especially for the highest energy emission, which is assumed to originate in the innermost part of the flow. 

This has clear implications for black hole spin estimates based on SED continuum fitting. In section 3.1 we demonstrate that standard disc model fits to the optical/UV spectra fail for the highest mass quasars such as J0927+0004 
when GR ray tracing is taken into account. This shows that the optical/UV spectrum is not simply a standard disc, so any spin estimate assuming standard disc models is not robust. 

For lower mass / higher Eddington ratio black holes, a (colour temperature corrected) disc model can provide adequate fits to the optical/UV data (\citealt{Mitchell23}) as the standard disc spectrum peaks in the EUV, so the impact of the GR ray tracing on the observable spectrum is small. However, it is also 
clear that extending the spectrum over a wider bandpass reveals non-disc emission, with the soft X-ray excess and high energy tail. We illustrate this using Fairall 9 in section 3.2, showing again that the highest energy emission is affected by GR ray tracing. 

This gives a potential way to constrain black hole spin from the energetics. However, we caution that there are caveats to this even within the model framework. We use the relativistic ray tracing transfer functions assuming that the radiation is emitted from a thin disc. In our model this is strictly speaking only true for the outer standard disc. The warm Comptonisation region and especially the hot corona may have a different geometry and velocity field, which will change the ray tracing, and may even affect the assumed emissivity.

Nevertheless, our new model still provides an improvement over older non-relativistic versions. Specifically, a non-relativistic version will overestimate the flux seen by an observer at low inclination, and hence could underestimate the power output of the AGN; specifically the mass accretion rate. We highlight this in our fit of Fairall 9, where we see a significant increase in the predicted mass accretion rate when relativistic effects are included, compared to the non-relativistic case.

\section*{Acknowledgements}
We would like to thank the anonymous referee for their helpful comments, which improved the manuscript.
We would also like to thank Delphine Porquet for using and testing early versions of the code, pointing out the occasional bug, and Vicky Fawcett for helpful conversations regarding the X-shooter data, and for providing the X-shooter spectrum of J0927+0004. 
Thank you also to Chris Belczynski and Jean-Pierre Lasota for motivating us to write this code.
SH acknowledges support from the Science and Technology Facilities Council (STFC) through the studentship grant ST/V506643/1. CD acknowledges the Science and Technology
Facilities Council (STFC) through grant ST/T000244/1 for support.

The {\sc python} version of the code uses the following moduels: {\sc scipy} \citep{Virtanen20}, {\sc numpy} \citep{Harris20}, and {\sc astropy} \citep{Astropy13, Astropy18, Astropy22}. Additionally, all plots were made using {\sc matplotlib} \citep{Hunter07}.

\section*{Data Availability}

The model code is publicly available through the corresponding authors GitHub: \url{https://github.com/scotthgn/RELAGN} (or through email upon request). 

The X-shooter data on J0927+0004, used in section 3.1, were obtained through private communication with V. Fawcett. They are also available through the ESO archive: \url{https://archive.eso.org/scienceportal/home}.
The Swift-UVOT data used in section 3.2 are available through the Swift archive: \url{https://www.swift.ac.uk/swift_live/index.php}, while the XMM-Newton data are archival, and can be directly accessed through HEASARCH: \url{https://heasarc.gsfc.nasa.gov/db-perl/W3Browse/w3browse.pl}



\bibliographystyle{mnras}
\bibliography{Refs} 




\appendix

\section{{\sc relagn} Model Documentation}

We have released two versions of the model code: one written in {\sc fortran} and the other in {\sc python}. The {\sc fortran} version is written to be used with {\sc xspec}, and so is specifically aimed at applying the model to spectral data. The {\sc python} version exists to provide more flexibility to the user, and is intended to provide more functionality to users who are mainly interested in spectral modelling. However, the default grids used in the calculations and the input parameters are identical between the two versions. These are described here.

The radial binning is set at 30 bins per decade (i.e $d\log r = 1/30$). This is chosen such that the radial resolution is fine enough that our calculations of the emission from each bin is accurate and that the relativistic effects are roughly constant across each bin. However, it is not so fine as to give numerical issues when applying the transfer functions.

Our code also defines its own energy grid when performing the calculations. This is a feature designed to ensure sufficient spectral coverage in the calculations when the code is being applied to data in {\sc xspec}; as {\sc xspec} will by default pass the data energy bins to the model subroutine. It also exists in the {\sc python} version, mostly as a convenience to the user. By default, the energy grid will extend from $E_{\text{min}} =  10^{-4}$\,keV to $E_{\text{max}} = 10^{3}$\,keV, using 2000 geometrically spaced bins. However, if {\sc xspec} were to pass an energy grid that extends beyond these limits, then the code will automatically adapt and re-scale its internal energy grid. In the {\sc python} version there exists a method for the user to manually re-scale the energy grid if needed.

Both versions of the code take the same input parameters, which are described in table \ref{tab:rel_pars}. There exists detailed documentation and example usage in the code repository (\url{https://github.com/scotthgn/RELAGN}).

{\renewcommand{\arraystretch}{1.6}
\begin{table}
    \centering
    \begin{tabular}{c|p{4cm}|c}
         Parameter & Description & Units \\
         \hline
         Mass & The black hole mass & $M_{\odot}$  \\
         Distance & The Co-Moving distance to the object & Mpc \\
         $\log \dot{m}$ & Mass accretion rate & $\dot{M}/\dot{M}_{\text{edd}}$ \\
         Spin & Black hole spin & Dimensionless \\
         $\cos (i)$ & Cosine of the inclination angle, measured from the z-axis (with the disc in the x-y plane) & Dimensionless \\
         $kT_{e, h}$ & Electron temperature for the hot Comptonisation component. For the {\sc fortran} version, if this is negative then ONLY the hot Comptonisation component is returned & keV \\
         $kT_{e, w}$ & Electron temperature for the warm Comptonisation component. For the {\sc fortran} version, if this is negative then ONLY the warm Comptonisation component is returned  & keV \\
         $\Gamma_{h}$ & Spectral index for hot Comptonisation component & Dimensionless \\
         $\Gamma_{w}$ & Spectral index for warm Comptonisation component. For the {\sc fortran} version, if this is negative then ONLY the standard disc component is returned & Dimensionless \\
         $r_{h}$ & Outer radius of the hot Comptonising region. If this is negative or less than $\risco$, then the code will default to $r_{\text{isco}}$ & $R_{G}$ \\
         $r_{w}$ & Outer radius of the warm Comptonising region. If this is negative or less than $\risco$, then the code will default to $r_{\text{isco}}$ & $R_{G}$ \\
         $\log r_{\text{out}}$ & Outermost disc radius. If this is negative, then the code will use the self-gravity radius $r_{sg}$ from \citet{Laor89} & $R_{G}$ \\
         $f_{\text{col}}$ & Colour-temperature correction. Note that this will ONLY be applied to the standard disc region. If this is negative, then the code will follow the relation given in \citet{Done12}. Otherwise it is treated as a constant correction across the standard disc region. & Dimensionless \\
         $h_{\text{max}}$ & Scale height of the hot Comptonisation region. Note that this is meant as a fine-tuning parameter. It will only affect the seed photon contribution to the hot Comptonisation luminosity. The code is also hardwired such that $h_{\text{max}} \leq r_{h}$. If $h_{\text{max}} > r_{h}$, then it will automatically re-set it such that $h_{\text{max}} = r_{h}$ & $R_{G}$ \\
         Redshift & Redshift of the source & Dimensionless
    \end{tabular}
    \caption{Parameters in {\sc relagn}. These are listed in the order they should be passed to the code.}
    \label{tab:rel_pars}
\end{table}
}

\section{Bonus model: {\sc relqso}}
As well as updating the {\sc agnsed} model, we have also applied the same updates to the {\sc qsosed} model. This is detailed in \citetalias{Kubota18}, and is a simplified version of {\sc agnsed}. To that extent, the method for applying the relativistic transfer functions is identical to that in {\sc relagn}; giving the model {\sc relqso}. However, there are a few important notes to be aware of before using {\sc relqso}.

A key concept in {\sc qsosed} is that hot Compton coronal emission is hardwired to $L_{h, \mathrm{diss}} = 0.02 L_{\mathrm{Edd}}$, which then sets the outer radius of the hot Comptonisation region $r_{h}$. When making the model relativistic, however, we have the question of whether this is the power emitted in the rest frame or as seen by the distant observer. For simplicity, we set the constraint that $L_{h, \mathrm{diss}} = 0.02 L_{\mathrm{Edd}}$ in the rest frame. Hence, in {\sc relqso} although the intrinsic X-ray power is always the same for any version of the model, when seen by a observer this is no longer true as now the apparent X-ray luminosity will be subject to full general relativistic effects.

{\sc qsosed} also self-consistently calculates the spectral index for the hot Compton region, $\Gamma_{h}$. Seed photons incident on the corona are up-scattered to higher energies, which in turn cools the corona. Hence, an increase in seed-photon power, $L_{h, \mathrm{seed}}$ relative to the dissipated power $L_{h, \mathrm{diss}}$ will give an increase in Compton-cooling, which in turn will give a softer spectrum (i.e $\Gamma_{h}$ increases, e.g \citealt{Beloborodov99}). It is then possible to estimate $\Gamma_{h}$ based off our values of $L_{h, \mathrm{diss}}$ and $L_{h, \mathrm{seed}}$, which we know as they are explicitly calculated within the model code. From \citetalias{Kubota18}, we can write:

\begin{equation}
    \Gamma_{h} = \frac{7}{3} \left( \frac{L_{h, \mathrm{diss}}}{L_{h, \mathrm{seed}}} \right)^{-0.1} 
\end{equation}

In {\sc relqso} when we calculate $\Gamma_{h}$ we use the values of $L_{h, \mathrm{diss}}$ and $L_{h, \mathrm{seed}}$ evaluated in the rest frame. As well as making the calculations simpler, the seed-photon and dissipated power seen by the hot corona do not depend on the position of the observer.

Additionally, in {\sc relqso} we set: $\kth = 100$\,keV, $\ktw = 0.2$\,keV, $\Gamma_{w} = 2.5$, $r_{w} = 2r_{h}$, and $h_{\mathrm{max}} = \mathrm{min}(10, r_{h})$. The remaining input parameters are listed in table \ref{tab:relqso_par}.

{\renewcommand{\arraystretch}{1.6}
\begin{table}
    \centering
    \begin{tabular}{c|p{4cm}|c}
         Parameter & Description & Units \\
         \hline
         Mass & The black hole mass & $M_{\odot}$  \\
         Distance & The Co-Moving distance to the object & Mpc \\
         $\log \dot{m}$ & Mass accretion rate & $\dot{M}/\dot{M}_{\text{edd}}$ \\
         Spin & Black hole spin & Dimensionless \\
         $\cos (i)$ & Cosine of the inclination angle, measured from the z-axis (with the disc in the x-y plane) & Dimensionless \\
         $f_{\mathrm{col}}$ & Colour-temperature correction. Note that this is ONLY applied to the standard disc region. & Dimensionless \\
         Redshift & Redshift of the source & Dimensionless \\
    \end{tabular}
    \caption{Parameters for {\sc relqso}.}
    \label{tab:relqso_par}
\end{table}
}

\section{The coronal solid angle}

A subtle point when calculating the emission from the hot Comptonised region is the seed photon power, $L_{h, \mathrm{seed}}$, seen by the corona. This depends on the solid angle subtended by the corona as seen from each annulus on the disc. Throughout we have followed \citetalias{Kubota18}, in order to be consistent with {\sc agnsed}, where they give the radially dependent covering fraction of the corona as:

\begin{equation}
    \frac{\Theta(R)}{2\pi} = \theta_{0} - \frac{1}{2} \sin(2\theta_{0})
\end{equation}

where $\sin(\theta_{0}) = H/R$. However, this is strictly speaking only correct in the two-dimensional case, whereas black hole accretion flows are three-dimensional.

For completeness we give below the derivation of the solid angle for both the 2D and 3D cases, in order to highlight the overall small difference this choice makes.

\subsection{The 2D solid angle}

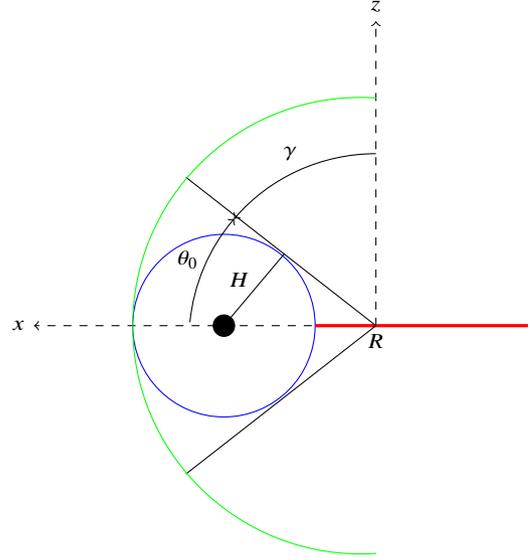
\begin{figure}
    \centering
    \begin{tikzpicture}

        \filldraw[color=black] (0, 0) coordinate (BH) circle (4pt) ;
        \draw[color=blue] (0, 0) circle (1.2) ; 

        \draw[->, dashed] (2, 0) coordinate (R) -- (-2.5, 0) coordinate (X) ;
        \draw[->, dashed] (2, 0) -- (2, 4) coordinate (Z);
        \node (zax) at (2, 4.2) {$z$} ;
        \node (xax) at (-2.7, 0) {$x$} ;

        \draw (0, 0) -- (0.8, 0.95) ;
        \node (H) at (0.2, 0.6) {$H$} ;

        \draw[very thick, color=red] (1.2, 0) -- (4, 0) ;
        \draw[color=black] (2, 0) -- (-0.5, 1.95) coordinate (P) ;
        \draw[color=black] (2, 0) -- (-0.5, -1.95) ;
        \node (R) at (2, -0.2) {$R$} ;

        \pic[draw, <-, "$\theta_{0}$", angle eccentricity=1.1, angle radius=70] {angle=P--R--BH} ;

        \pic[draw, ->, "$\gamma$", angle eccentricity=1.1, angle radius=70] {angle=Z--R--P} ;
        \clip (-2, -3) rectangle (2, 3) ;
        \draw[color=green] (1.8, 0) circle (3) ;


    \end{tikzpicture}
    \caption{Schematic of the disc-corona geometry used to derive the corona covering fraction. The red line indicates the disc, while the blue circle shows the corona. The green semi-circle indicates the projected sky, used to determine the covering-fraction. $H$ is the scale-height of the corona, and $R$ is the disc radius being considered.}
    \label{fig:corona_solid}
\end{figure}

In 2 dimensions the visibility of corona, as seen from the disc, is:

\begin{equation}
    \Theta = 2 \int_{0}^{\theta_{0}} \sin(\theta) \frac{\cos(\gamma)}{0.5} d\theta
\end{equation}

where the factor 2 comes from the disc having two sides, $\cos(\gamma)/0.5$ is the disc visibility at radius $R$ for an observer at inclination $\gamma$, and $\sin(\theta)$ comes from the definition of the solid angle. 
Note that the factor 0.5 in the disc visibility comes from results of radiative transfer through a photosphere simulations that show the emission is more or less isotropic for inclinations $i < 60$\,deg \citep{Davis11}. Hence we normalise all disc emission by $\cos(i)/\cos(60)$, as also done in \citet{Done13} and \citet{Kubota18}.
We also stress here that $\theta$ is measured from the x-axis up towards the z-axis, unlike the standard definitions. This can be thought of as simply rotating the standard coordinate system (where $\theta$ is measured from the z-axis) by $\pi/2$, and is chosen because it makes the derivation simpler.

From Fig.\,\ref{fig:corona_solid} we see that $\gamma = \pi/2 - \theta$, and so we can write $\cos(\gamma) = \sin(\theta)$. Eqn. C2 now becomes:

\begin{equation}
    \Theta = 4\int_{0}^{\theta_{0}} \sin^{2}(\theta) d\theta = 4\int_{0}^{\theta_{0}} \left( \frac{1}{2} - \frac{1}{2} \cos(2\theta) \right) d\theta
\end{equation}

Solving the above integral, and dividing by $2\pi$, we arrive at Eqn. C1.

\begin{figure*}
    \centering
    \includegraphics[width=\textwidth]{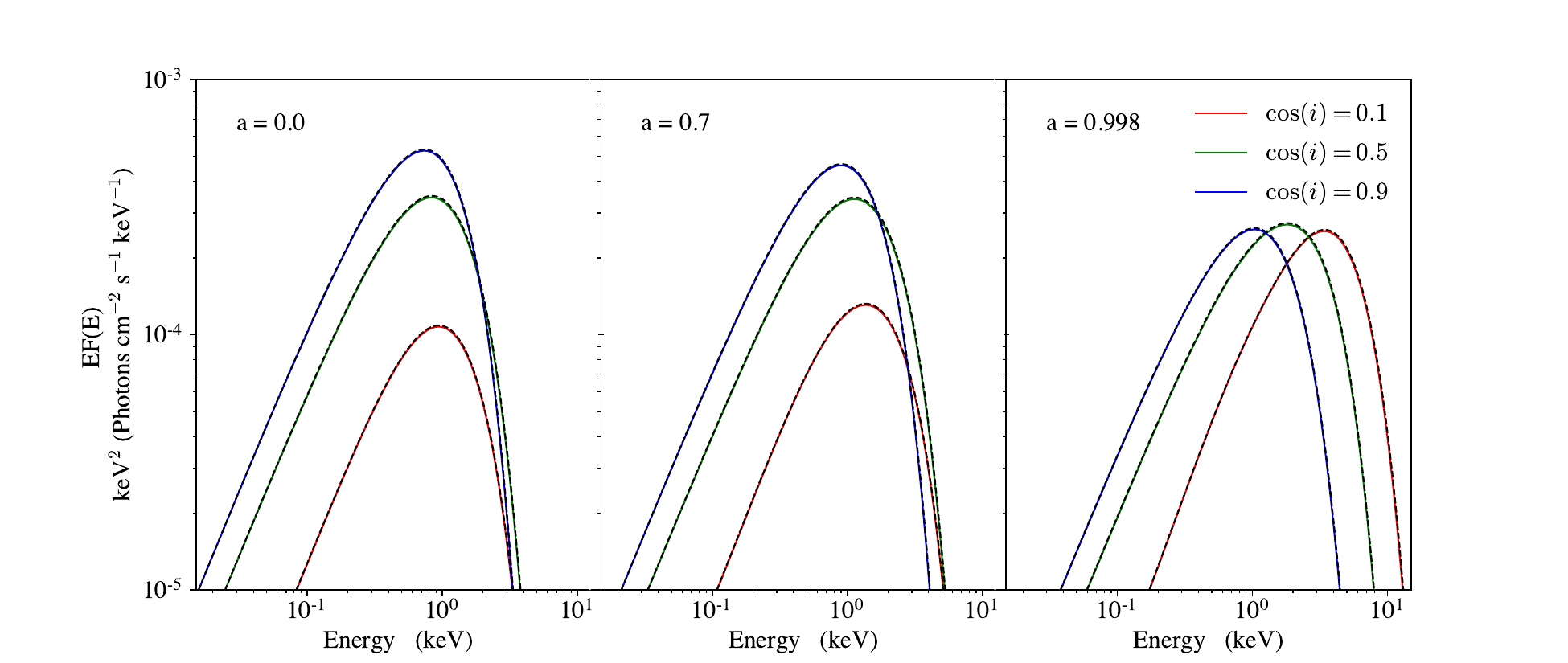}
    \caption{Comparison of {\sc relagn} to {\sc kerrbb} for a $\Mbh = 10 \Msol$ black hole, accreting at $\log \mdot = -1$. The solid coloured lines show the {\sc relagn} model for a given spin and inclination, while the dashed black lines show {\sc kerrbb} for the same parameters.}
    \label{fig:comp2kerrbb}
\end{figure*}

\subsection{The 3D solid angle}

In 3 dimensions the solid angle of the corona, as seen by some point on the disc, is identical to that of a conic section. Hence, when including the disc visibility, we can write:

\begin{equation}
    \Omega = \int_{\phi=0}^{2\pi} \int_{\theta=0}^{\theta_{0}} \sin(\theta) \frac{\cos(\gamma)}{0.5} d\theta d\phi
\end{equation}

Note that since we are now integrating over all $\phi$, we no longer need to include the factor $2$ outside the integral. Clearly this is almost identical to Eqn. C2, hence we can solve it in the same manner. The resulting covering fraction, is then:

\begin{equation}
    \frac{\Omega}{4\pi} = \frac{1}{2} \left( \theta_{0} - \frac{1}{2} \sin(2\theta_{0}) \right)
\end{equation}

It is important to note here that this solid angle derivation is only an approximation of the real system solid angle. Firstly, we have neglected light-bending, which although will make a negligible impact for a large corona, can have a stronger impact if the corona size reduces sufficiently. Additionally, we have made no assumptions about the optical thickness of the corona. In reality it would be expected that the optical thickness should change with $\theta$, going from 0 at $\theta_{0}$ to a maximum along the equatorial plane. However, calculating this would require assumptions about the currently unconstrained radial and vertical density structure of the hot corona. Hence, for simplicity we make no assumptions about the coronal optical thickness.

\section{Comparison to {\sc kerrbb}}

We test our implementation of the relativistic transfer functions by comparison to {\sc kerrbb} \citep{Li05}, which calculates the continuum disc emission including all relativistic corrections. {\sc kerrbb} only considers thermal (or colour temperature corrected thermal) emission from a standard disc, extending from $\risco$ to infinity. Hence, to compare to our {\sc relagn} code we set $r_{w} = r_{h} = \risco$, and set $\rout = 10^{7} R_{G}$. Fig.\,\ref{fig:comp2kerrbb} shows this comparison for a range of black hole spin and inclination. It can be seen that our {\sc relagn} model closely follows {\sc kerrbb} at all spins and inclinations.


\bsp	
\label{lastpage}
\end{document}